\documentclass[twocolumn,aps, superscriptaddress]{revtex4}
\usepackage{amsmath}
\usepackage{amssymb}
\usepackage{graphicx}
\usepackage{color}

\newcommand{\be}{\begin{equation}}
\newcommand{\ee}{\end{equation}}
\newcommand{\bea}{\begin{eqnarray}}
\newcommand{\eea}{\end{eqnarray}}
\newcommand{\bse}{\begin{subequations}}
\newcommand{\ese}{\end{subequations}}

\newcommand{\tcs}{${\rm ThCr_2Si_2}$}
\newcommand{\cca} {CaCo$_{2-y}$As$_2$}
\newcommand{\sca} {${\rm SrCo_2As_2}$}
\newcommand{\bfa} {${\rm BaFe_2As_2}$}
\newcommand{\bca} {${\rm BaCo_2As_2}$}

\newcommand{\csca}{Ca$_{1-x}$Sr$_x$Co$_{2-y}$As$_2$}
\newcommand{\ecna}{Eu(Co$_{1-x}$Ni$_x$)$_{2-y}$As$_2$}
\newcommand{\scna}{Sr(Co$_{1-x}$Ni$_x$)$_2$As$_2$}
\newcommand{\ccia}{Ca(Co$_{1-x}$Ir$_x$)$_{2-y}$As$_2$}
\newcommand{\bcia}{Ba(Co$_{1-x}$Ir$_x$)$_2$As$_2$}

\begin{document}

\title{Emergence of ferromagnetism due to Ir substitutions in single-crystalline Ba(Co$_{1-x}$Ir$_x$)$_2$As$_2$ ($0 \leq x \leq 0.25$)}

\author{Santanu Pakhira}
\affiliation{Ames Laboratory, Iowa State University, Ames, Iowa 50011, USA}
\author{N. S. Sangeetha}
\affiliation{Ames Laboratory, Iowa State University, Ames, Iowa 50011, USA}
\author{V. Smetana}
\affiliation{Department of Materials and Environmental Chemistry, Stockholm University, Svante Arrhenius v\"{a}g 16 C, 106 91 Stockholm, Sweden}
\author{A.-V. Mudring}
\affiliation{Department of Materials and Environmental Chemistry, Stockholm University, Svante Arrhenius v\"{a}g 16 C, 106 91 Stockholm, Sweden}
\author{D. C. Johnston}
\affiliation{Ames Laboratory, Iowa State University, Ames, Iowa 50011, USA}
\affiliation{Department of Physics and Astronomy, Iowa State University, Ames, Iowa 50011}

\date{\today}

\begin{abstract}

The ternary-arsenide compound \bca\ was previously proposed to be in proximity to a quantum-critical point where long-range ferromagnetic (FM) order is suppressed by quantum fluctuations. Here we report the effect of Ir substitution for Co on the magnetic and thermal properties of \bcia\ ($0 \leq x \leq 0.25$) single crystals. These compositions all crystallize in an uncollapsed body-centered-tetragonal ${\rm ThCr_2Si_2}$ structure with space group $I4/mmm$. Magnetic susceptibility $\chi$ measurements reveal clear signatures of FM ordering for $x \geq 0.11$ with a nearly composition-independent Curie temperature $T_{\rm C} \approx 13$~K\@.  The small variation of $T_{\rm C}$ with~$x$, the occurrence of hysteresis in magnetization versus field isotherms at low field and temperature, very small spontaneous and remanent magnetizations $<0.01\,\mu_{\rm B}$/f.u., and thermomagnetic irreversibility in the low-temperature region together indicate that the FM response arises from short-range FM ordering of spin clusters as previously inferred to occur in Ca(Co$_{1-x}$Ir$_x$)$_{2-y}$As$_2$.  Heat-capacity $C_{\rm p}(T)$ data do not exhibit any clear feature around $T_{\rm C}$, further indicating that the FM ordering is short-range and/or associated with itinerant moments. The $C_{\rm p}(T)$ in the paramagnetic temperature regime 25--300~K is well described by the sum of a Sommerfeld electronic contribution and Debye and Einstein lattice contributions where the latter suggests the occurrence of low-frequency optic modes associated with the heavy Ba atoms in the crystals.

\end{abstract}

\maketitle

\section{Introduction}
Since the discovery of high-temperature superconductivity (SC) at a transition temperature $T_{\rm c}$ = 38 K in K-doped \bfa\ ~\cite{Rotter2008}, ternary pnictides having general formula $AM_2X_2$ ($A$ = Ca, Sr, Ba, Eu, $M$ = transition metal, and $X$ = As, Sb, P) have become the playground to explore different novel states governed by complex interplay between superconductivity and magnetism~\cite{Johnston2010, Canfield2010, Paglione2010, Fernandes2010, Stewart2011, Scalapino2012, Dai2012, Dagotto2013, Dai2015, Si2016}. Numerous investigations on a large number of such systems have been carried out to understand the unconventional SC in Fe-based arsenide compounds. On the other hand, although Co-based pnictide compounds do not exhibit SC, various novel magnetic states associated with itinerant Co moments have been discovered that are rich in physics~\cite{Quirinale2013, Anand2014Ca, Pandey2013, Jayasekara2013, Sefat2009, Anand2014, Sangeetha_EuCo2P2_2016, Sangeetha_EuCo2As2_2018}. Most of these 122-type compounds crystallize in the body-centered tetragonal \tcs\ structure with space group $I4/mmm$. In this structure, the transition-metal atoms form a square lattice and are tetrahedrally coordinated by the main-group elements. The electronic band structure and associated Fermi surface for this structure are sensitive to perturbations such as external pressure, temperature, and most importantly, chemical doping.

Tuning of the physical properties of these 122-type materials through chemical doping has paved the way for significant discoveries in the field of superconductivity and magnetism. As examples, although $A$Fe$_2$As$_2$ ($A$ = Sr, Ca, and Eu) exhibit long-range AFM order, superconductivity can be induced by both electron and hole doping in these systems~\cite{Sasmal2008, Jasper2008, Kumar2009, Ren2009}. In the absence of long-range magnetic order, strong dynamic short-range AFM correlations act as a glue for the cooper-pair formation required for SC to appear in these systems~\cite{Johnston2010, Canfield2010, Paglione2010, Fernandes2010, Stewart2011, Scalapino2012, Dai2012, Dagotto2013,Dai2015}. Significant changes in the superconducting and magnetic properties have been reported both theoretically and experimentally by partially substituting the Fe atoms in $A$Fe$_2$As$_2$ by the transition metals Cr, Mn, Co, Ni, Cu, Ru, Rh, and Pd~\cite{Sefat2009prb_bfca, Marty2011prb_bfca, Kasinathan2009njp, Kim2010prb, Sefat2008prl, Canfield2009prb, Liu2011prb, Ideta2013prl, Thaler2010prb, Dhaka2011prl, Ni2009prb, Han2009prb}. A small amount (1.6\%) of K doping in the antiferromagnetic (AFM) insulator BaMn$_2$As$_2$ gives rise to a metallic state in Ba$_{1-x}$K$_x$Mn$_2$As$_2$~\cite{Pandey2012prl}.

Among the CoAs-based systems, \cca\ forms in the collapsed-tetragonal (cT) structure~\cite{Anand2012} with vacancies in the Co site and exhibits A-type AFM order with N\'eel  temperature $T_{\rm N} = 52$--77~K, depending on the crystal~\cite{Anand2014Ca, Cheng2012, Ying2012}. Inelastic neutron-scattering measurements further reveal that \cca\ is a unique itinerant magnetic system where strong magnetic frustration is present within the $J_1$-$J_2$ Heisenberg model on a square lattice~\cite{Sapkota2017}. The frustration parameter is found to be $\eta = J_1/(2J_2)=-1.03(2)$ with nearest-neighbor ferromagnetic (FM) interactions ($J_1 < 0$).

The compound \sca\ crystallizes in the uncollapsed-tetragonal (ucT) structure~\cite{Anand2012}.  The paramagnetic (PM) state is Stoner-enhanced~\cite{Pandey2013}. Although no magnetic ordering or SC is observed above 0.05~K~\cite{Li2019}, stripe-type AFM fluctuations and competing magnetic phases have been evidenced in \sca\ via inelastic neutron-scattering studies~\cite{Jayasekara2013,Li2019}.

The intermixing of Ca and Sr in \csca\ results in an anomalous composition-induced crossover in the AFM ordering direction between the $c$~axis and $ab$~plane~\cite{Ying2013,Sangeetha2017}. The composition with $x = 0.33$ is of special interest because it shows a temperature-induced transition between these ordering directions~\cite{Sangeetha2017}, a rarely-observed phenomenon. Magnetic neutron-diffraction measurements~\cite{Li2019csca} revealed that the AFM propagation vector is $\vec{\tau} = (0,0,1)$ (A-type AFM, $+-+-$) for $0\leq x \lesssim0.2$ with the ordered moments $\vec{\mu}\parallel c$, $\vec{\tau} = (0,0,1/2)$ ($++--$) for $0.2\lesssim x \lesssim0.3$ with $\vec{\mu}\parallel c$, and $\vec{\tau} = (0,0,1/2)$ ($++--$) or a 90$^\circ$ $c$-axis helix with $\vec{\mu}\perp c$ for $0.3\lesssim x \lesssim 0.5$.  The material is PM for $0.5 \lesssim x \leq 1$.  The ordered-moment orientations for $0\leq x \lesssim0.2$ and $0.3\lesssim x \lesssim 0.5$ are the same as deduced from the magnetic susceptibility $\chi$ measurements~\cite{Ying2013, Sangeetha2017}; however, these $\chi$ measurements gave no information about the nature of the magnetic ordering for $0.2\lesssim x \lesssim 0.3$.

The Eu$^{+2}$ spins~7/2 in EuCo$_{2-y}$As$_2$ order below $T_{\rm N} \approx 47$~K in a $c$-axis helical AFM structure with a turn angle $\approx 0.80\pi$~\cite{Sangeetha_EuCo2As2_2018, Tan2016, Reehuis1992}. The neutron-diffraction measurements indicated that the Co atoms do not contribute to the ordered moment~\cite{Reehuis1992}.  Enhanced ordered and effective moments of the Eu spins are observed from magnetization versus applied magnetic field $M(H)$ isotherms and $\chi(T)$ studies~\cite{Sangeetha_EuCo2As2_2018}.

Chemical substitutions on the Co site have also been found to result in intriguing changes in the magnetic ordering in $A$Co$_2$As$_2$ systems. The A-type AFM order in \cca\ is smoothly suppressed by hole doping in Ca(Co$_{1-x}$Fe$_x$)$_{2-y}$As$_2$ and vanishes for $x > 0.12$~\cite{Jayasekara2017prb}. A smooth crossover between the cT phase for $x = 0$ and ucT phase for $x = 1$ is also observed with increasing $x$ from x-ray diffraction studies, with a corresponding crossover in the magnetic properties~\cite{Jayasekara2017prb}.  A composition-induced quantum-critical point associated with non-Fermi-liquid behavior was discovered in \scna\ crystals at $x \approx 0.3$~\cite{Sangeetha2019scna}.  Below the critical concentration $x \approx 0.3$, the crystals exhibit $c$-axis helical AFM ordering~\cite{YLi2019, Sangeetha2019scna, Wilde2019}. Single crystals of \ecna\ exhibit a complex magnetic phase diagram in the magnetic field~$H$-temperature~$T$ plane~\cite{Sangeetha2020ecna}. Although the end members $x = 0$ and 1 both exhibit $c$-axis helical AFM ordering, magnetization data for $x = 0.03$ and 0.10 indicate the possible presence of a 2$q$ magnetic structure containing two helix axes along the $c$ axis and in the $ab$ plane, respectively. Along with the AFM ordering of the localized Eu$^{+2}$ spins, coexistence of itinerant FM ordering associated with the Co/Ni atoms was also evident for $0.20 \leq x \leq 0.65$.

Metallic \bca\ crystallizes in a ucT structure~\cite{Anand2012}. Although the compound does not exhibit long-range magnetic ordering or SC down to 1.8~K, the Stoner-enhanced $\chi$ of this PM compound is quite large and exhibits an upturn in the low-temperature region~\cite{Sefat2009, Anand2014}. The $\chi(T)$ data also show a broad maximum around $\sim150$~K for crystals grown out of {\mbox{Co-As} self-flux~\cite{Sefat2009}. A large density of states at the Fermi energy ${\cal D}(E_{\rm F}) \approx 18$~states/eV~f.u.\ obtained from the Sommerfeld coefficient~$\gamma$ via analysis of low-$T$ heat capacity data is associated with the strong electron-electron correlations and the electron-phonon interaction present in the system and the value is larger than estimated from band-structure calculations [${\cal D}(E_{\rm F}) = 8.23$~states/eV~f.u.].  ARPES data and band-structure analysis demonstrate the absence of Fermi-surface nesting in this compound in contrast to that observed in the Fe-based analog~\cite{Dhaka2013}.

Based on the experimental results and theoretical calculations it was suggested that \bca\ is in proximity to a quantum-critical point where long-range FM order is suppressed by quantum fluctuations~\cite{Sefat2009}. The proximity to a quantum-critical point is generally expected to be able to be tuned by different perturbations such as chemical doping and/or external pressure. A high-pressure study revealed that the substantial electron correlations in \bca\ are reduced with increasing pressure up to 8~GPa~\cite{Ganguli2013}, with no evidence for magnetic ordering.  A small amount of K doping on the Ba site with composition Ba$_{0.94}$K$_{0.06}$Co$_2$As$_2$ results in weak ferromagnetism at $T \lesssim 10$~K with ordered moments of 0.007 and 0.03 $\mu_{\rm B}$/f.u.\ for two batches of self-flux-grown crystals~\cite{Anand2014}.

Although numerous studies have been carried out on the 122-type pnictide compounds with mainly 3$d$ and in some cases with 4$d$ elements~\cite{Johnston2010, Canfield2010, Paglione2010, Stewart2011}, those containing 5$d$ transition elements such as Ir are rare.   Other Ir-based compounds such as ${\rm Sr_2IrO_4}$ have attracted much attention due to the presence of strong spin-orbit coupling~\cite{Crawford1994, Bertinshaw2019}.  Recently, we reported that with increasing Ir-substitution,  \ccia\ single crystals exhibit a composition-induced crossover from A-type AFM to a FM cluster-glass phase~\cite{Pakhira2020ccia}. In the present work, we report the effect of Ir substitution on the Co site in \bca\ single crystals including similar evidence for the occurrence of a FM cluster-glass phase.

The experimental details are given in Sec.~\ref{Sec:ExpDet}.  The crystallography results and analyses are presented in Sec.~\ref{Sec:Cryst}, magnetization data in Sec.~\ref{Sec:MagData}, and heat capacity data in Sec.~\ref{Sec:Cp}.  Concluding remarks are given in Sec.~\ref{Sec:Summary}.

\section{\label{Sec:ExpDet} Experimental details}

Single crystals of \bcia\ with $x$ = 0, 0.03, 0.07, 0.11, 0.18, and 0.25 were grown out of self flux in a molar ratio Ba:Co:Ir:As = 1:4($1 - x$):4$x$:4 using the high-temperature solution-growth technique. The high-purity starting materials Ba (99.999\%, Alfa Aesar), Co (99.998\%, Alfa Aesar), Ir (99.9999\%, Ames Laboratory), and As (99.9999\%, Alfa Aesar) were placed in an alumina crucible. The Ba pieces were placed at the bottom of the crucible to avoid oxidation while transferring the crucible into a silica tube for sealing. A wad of quartz wool was placed slightly above the crucible to facilitate extraction of the excess flux during centrifugation. The crucible containing the starting materials was sealed in a silica tube under $\approx$ 1/4 atm of Ar gas. The sealed silica tube was then preheated at 650~$^{\circ}$C for 6~h and then heated to 1300~$^{\circ}$C for 20~h for homogenization. In both cases the  temperature ramp rate was 50~$^{\circ}$C/h. Finally, the sealed tube was cooled to 1180~$^{\circ}$C at a rate of 6~$^{\circ}$C/h and the single crystals were then separated by decanting the excess flux using a centrifuge. Plakelike single crystals with typical cross-sectional area $2\times2$~mm$^2$ to $1\times1$~mm$^2$ were obtained. The crystal size and homogeneity both decrease with increasing Ir concentration.  The crystals are sensitive to air as evidenced by the surfaces of the crystals turning black upon exposure to air.  Measurable crystals were not obtained for $x > 0.25$.

A JEOL scanning-electron microscope (SEM) equipped with an energy-dispersive x-ray spectroscopy  (EDS) attachment was employed to check the phase homogeneity of the obtained \bcia\ crystals. The average compositions of the crystals were determined by measuring compositions at different points of both the surfaces for each crystal using EDS and the most homogeneous crystals were used for the physical-property measurements. The average compositions along with their uncertainties for the crystals used for the physical-property  measurements are listed in Table~\ref{CrystalData} below.

\begin{figure}
\includegraphics[width = 3in]{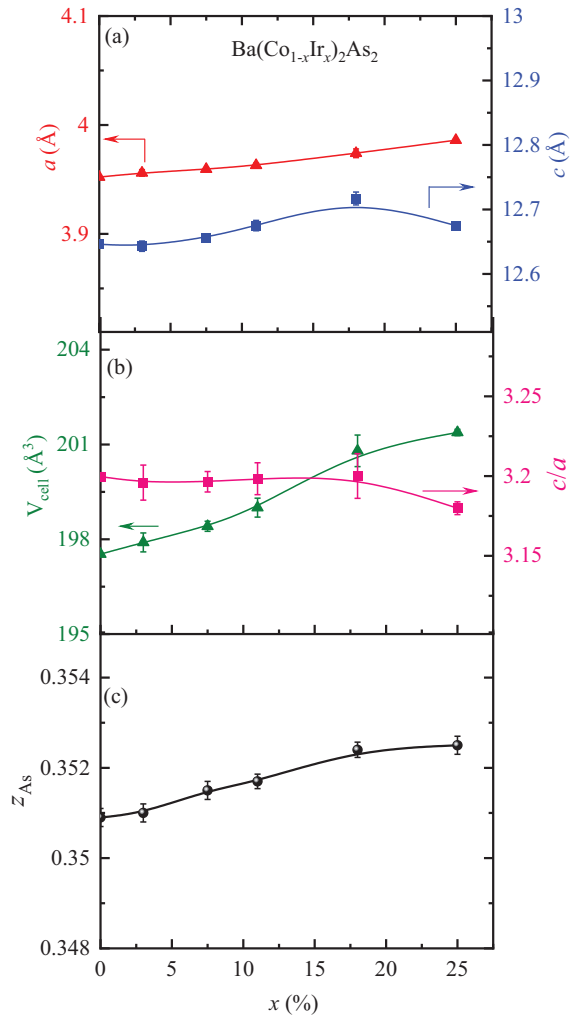}
\caption{Crystallographic parameters (a)~tetragonal lattice parameters $a$ and $c$, (b)~unit cell volume $V_{\rm cell}$ and $c/a$ ratio, and (c)~$c$-axis As position parameter $z_{\rm As}$ versus the Ir concentration~$x$ in the \bcia\ crystals.}
\label{unitcellparameters}
\end{figure}

Single-crystal x-ray diffraction (XRD) measurements were performed at room temperature on a Bruker D8 Venture diffractometer operating at 50~kV and 1~mA equipped with a Photon 100 CMOS detector, a flat graphite monochromator and a Mo~K$\alpha$ I$\mu$S microfocus source ($\lambda = 0.71073$~\AA). The raw frame data were collected using the Bruker APEX3 program \cite{APEX2015}, while the frames were integrated with the Bruker SAINT software package \cite{SAINT2015} using a narrow-frame algorithm for  integration of the data and were corrected for absorption effects using the multiscan method (SADABS) \cite{Krause2015}.  The atomic thermal factors were refined anisotropically.  Initial models of the crystal structures were first obtained with the program SHELXT-2014 \cite{Sheldrick2015A} and refined using the program SHELXL-2014 \cite{Sheldrick2015C} within the APEX3 software package.

Magnetization $M$ measurements as a function of~$T$ and applied magnetic field~$H$ were carried out using a Quantum Design, Inc., Magnetic-Properties Measurement System in the temperature range 1.8--300~K and with $H$ up to 5.5~T (1~T~$\equiv10^4$~Oe). The zero-field heat capacity $C_{\rm p}(T)$ measurements were performed using the relaxation technique in a Quantum Design, Inc., Physical-Properties Measurement System in the temperature range 1.8--300~K\@.

\section{\label{Sec:Cryst} Crystallography}

\begin{table*}
\caption{\label{CrystalData} Room-temperature crystallographic data for \bcia\ ($0 \leq x \leq 0.25$) single crystals. The labeled compositions were obtained from EDS analyses. The tetragonal lattice parameters $a$ and~$c$, unit cell volume $V_{\rm cell}$, $c/a$ ratio, and fractional $c$-axis position of the As site ($z_{\rm As}$) were obtained from the single-crystal XRD analyses.}
\begin{ruledtabular}
\begin{tabular}{ cccccc }
 Compound  & $a$ (\AA)  & $c$ (\AA) & $V_{\rm cell}$ (\AA$^3$) & $c/a$ & $z_{\rm As}$\\
\hline
BaCo$_2$As$_2$                               &   3.9522(4)   &   12.646(2)   &   197.53(4)   &   3.199(2)   &   0.3509(2)   \\
Ba(Co$_{0.97(1)}$Ir$_{0.03(1)}$)$_2$As$_2$    &   3.956(3)    &   12.643(8)    &   197.9(3)  &   3.196(1)   &   0.3510(2)   \\
Ba(Co$_{0.93(1)}$Ir$_{0.07(1)}$)$_2$As$_2$    &   3.959(1)    &   12.656(5)    &   198.4(2)  &   3.196(6)   &   0.3515(2)   \\
Ba(Co$_{0.89(1)}$Ir$_{0.11(1)}$)$_2$As$_2$      &   3.963(2)    &   12.675(8)    &   199.0(3)  &   3.198(10)   &   0.3517(16)   \\
Ba(Co$_{0.82(2)}$Ir$_{0.18(2)}$)$_2$As$_2$      &   3.974(4)    &   12.72(1)    &   200.8(5)  &   3.200(14)   &    0.3524(17)  \\
Ba(Co$_{0.75(2)}$Ir$_{0.25(2)}$)$_2$As$_2$      &   3.986(1)    &   12.675(3)    &   201.4(1)  &   3.180(4)   &   0.3525(2)   \\
\end{tabular}
\end{ruledtabular}
\end{table*}

The room-temperature single-crystal XRD measurements showed that all the \bcia\ crystals in Table~\ref{CrystalData} form in the body-centered tetragonal \tcs\ structure with space group $I{\rm 4}/mmm$. The tetragonal lattice parameters $a$ and~$c$, unit cell volume $V_{\rm cell}$, $c/a$ ratio, and fractional $c$-axis position coordinate of the As site ($z_{\rm As}$) are listed in Table~\ref{CrystalData} and plotted in Figs.~\ref{unitcellparameters}(a)--\ref{unitcellparameters}(c). Whereas $a$ increases monotonically and nearly linearly with Ir content $x$, $c(x)$ exhibits a nonmonotonic variation, resulting in a nonlinear variation of $V_{\rm cell}$ with~$x$. The $c/a$ ratio clearly indicates that the Ir-substituted crystals all form in the ucT structure~\cite{Anand2012}, as does the undoped \bca\ compound. The $z_{\rm As}$ value also increases slightly with $x$. The EDS measurements of the compositions of the crystals in Table~\ref{CrystalData} reveal that the homogeneity of the Ir concentration slightly decreases with increasing Ir concentration.

\section{\label{Sec:MagData} Magnetic measurements}

\subsection{Magnetic susceptibility}

\begin{figure*}[ht!]
\includegraphics[width = 5.in]{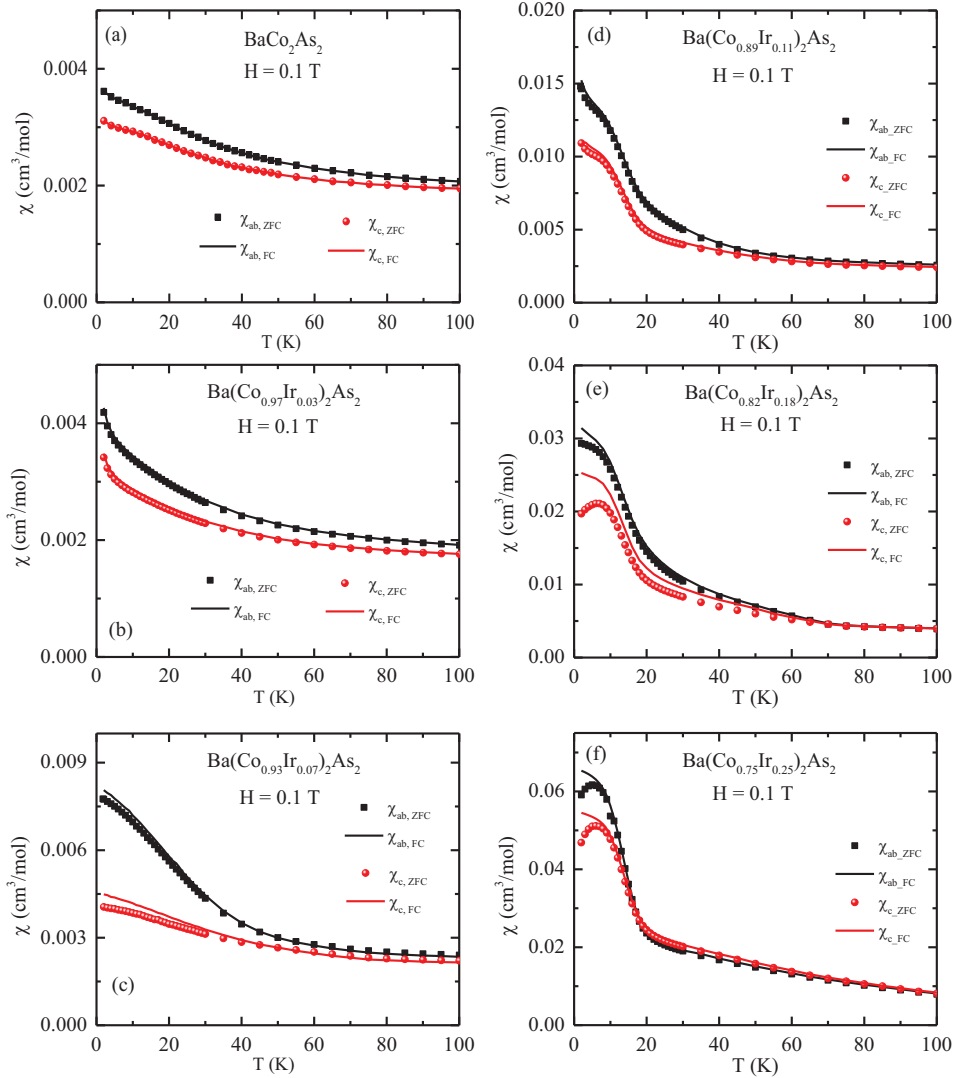}
\caption{Magnetic susceptibility ($\chi_\alpha = M_\alpha/H,\ \alpha = ab,c$) vs temperature~$T$ for \bcia\ crystals in a magnetic field $H =$ 0.1~T applied in the $ab$~plane ($H \parallel ab$) and along the $c$~axis ($H \parallel c$). Field-cooled (FC) and zero-field-cooled (ZFC) data are shown for each crystal and field direction.  Note the different scales for the ordinates of the plots.}
\label{M-T_all_separate}
\end{figure*}

\begin{figure}[h]
\includegraphics[width = 3.in]{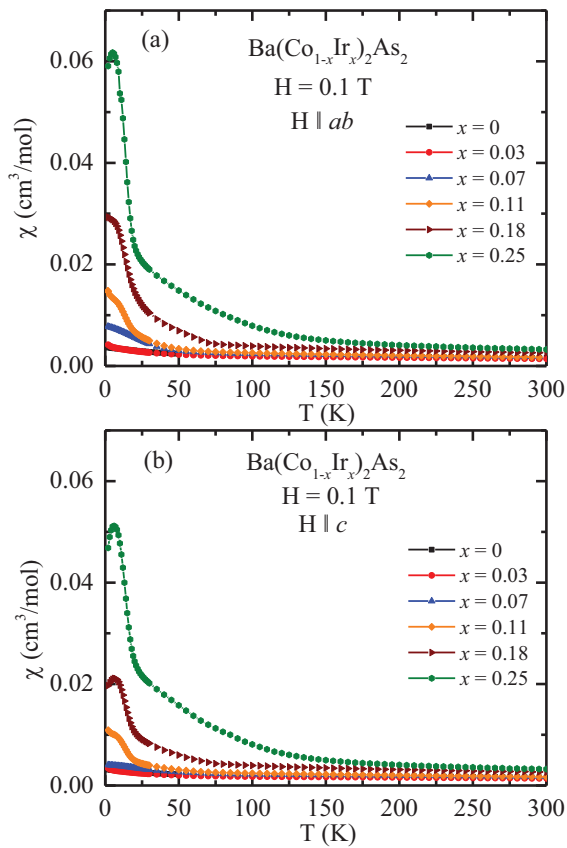}
\caption{Comparison of the temperature $T$-dependent ZFC magnetic susceptibility~$\chi$ of \bcia\ crystals for $H = 0.1$~T magnetic field applied (a)~in the $ab$~plane ($H \parallel ab$) and (b)~along the $c$~axis ($H \parallel c$).}
\label{M-T_all_together}
\end{figure}

\begin{figure}
\includegraphics[width = 2.74in]{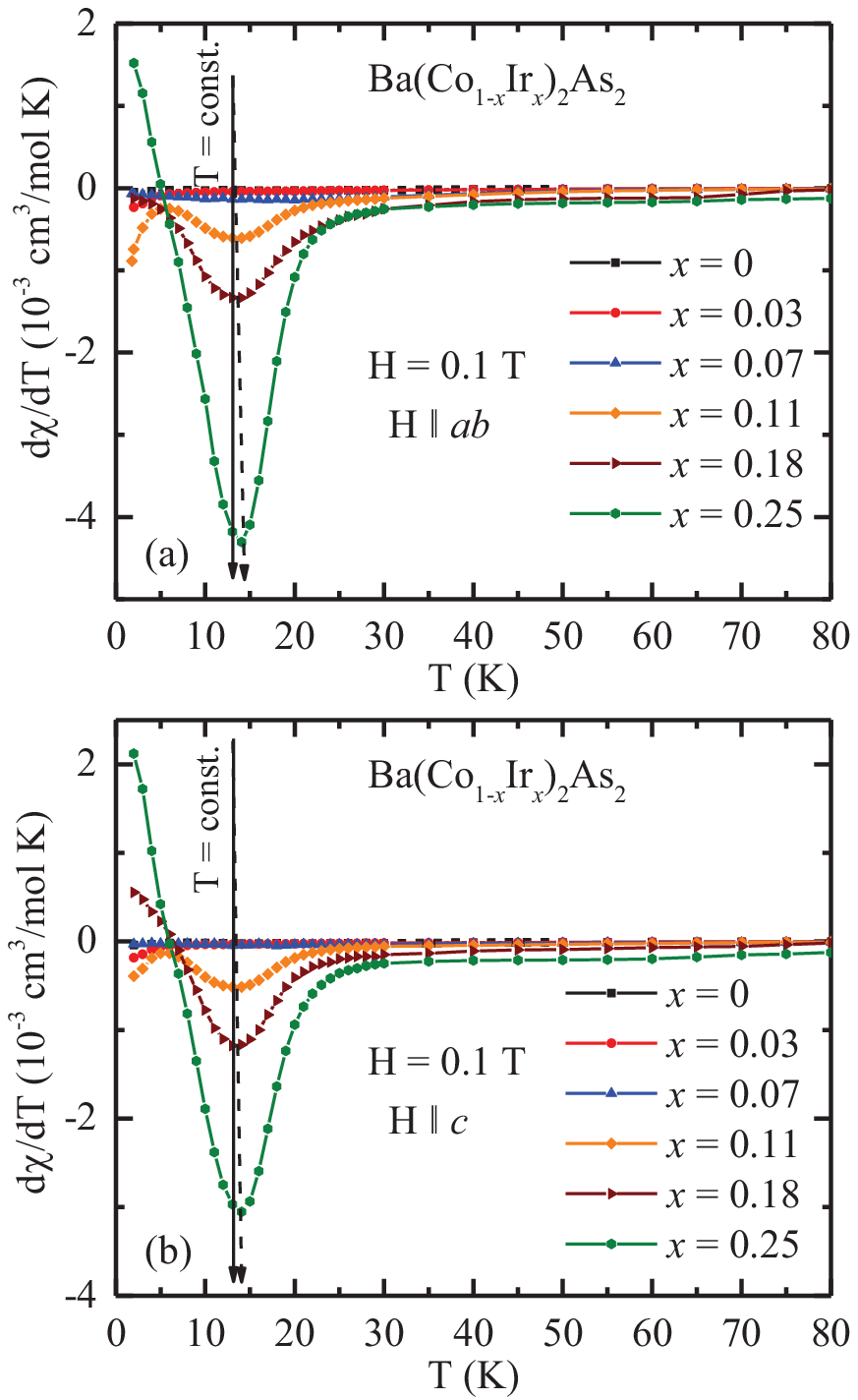}
\caption{Temperature derivative $d\chi(T)/dT$ of the zero-field-cooled (ZFC) magnetic susceptibility $\chi$ of \bcia\ crystals as a function of temperature~$T$ for $H =$ 0.1~T applied (a)~in the $ab$~plane ($H \parallel ab$) and (b)~along the $c$~axis ($H \parallel c$).  }
\label{dChi_dT_all_together}
\end{figure}

The $T$ dependences of the magnetic susceptibility $\chi = M/H$ for the \bcia\ crystals measured under both zero-field-cooled (ZFC) and field-cooled (FC) conditions in  $H = 0.1$~T applied in the $ab$~plane ($H~ \| ~ab$) and along the $c$~axis ($H~ \| ~c$) are shown in Figs.~\ref{M-T_all_separate}(a)--\ref{M-T_all_separate}(f). No magnetic ordering is observed down to 1.8~K for \bca, as reported earlier~\cite{Sefat2009,Anand2014}. All crystals exhibit increases in $\chi_\alpha(T)~(\alpha = ab,\ c)$ with decreasing~$T$ for $T \lesssim 50$~K compared with the higher-$T$ data. Both the nature and magnitude of the low-$T$ upturns change with increasing Ir substitution.  The maximum value of the FC $\chi_{ab}$ for each~$x$ occurs at $T=2$~K and increases from $\approx 0.0037~{\rm cm^3/mol}$ for $x=0$ to $\approx 0.065~{\rm cm^3/mol}$ for $x=0.25$.  Little bifurcation between the ZFC and FC susceptibilities is observed for $x=0,$  $x=0.03$ and 0.11, but bifurcation is significant for $x = 0.07$, 0.18, and 0.25.  Significant anisotropy is seen between $\chi_{ab}$ and $\chi_c$ for all crystals especially at the lower temperatures, with $\chi_{ab} > \chi_c$.

The temperature dependences of $\chi_\alpha$ ($\alpha=ab,\ c$) clearly change for $x\geq 0.07$, where the $\chi_{\alpha}(T)$ exhibit upturns and inflection points on cooling below $\sim 20$~K, together with the development of broad peaks at $\sim 6$--8~K for $x = 0.18$ and~0.25.  For these two compositions, the ZFC and FC data also show especially large bufurcations.  These data suggest the development of some type of FM ordering below $\sim 20$~K with increasing Ir concentration.

Comparisons of the ZFC $\chi_\alpha~(\alpha=ab~{\rm or}~c$) data for the \bcia\ crystals are shown in Figs.~\ref{M-T_all_together}(a) and~\ref{M-T_all_together}(b) for $H \parallel ab$ and $H \parallel c$, respectively.  These data for the two field directions again consistently show clear evidence for the onset of some type of FM order below the same temperature $T\sim 20$~K for $x=0.11$, 0.18, and 0.25.

A  quantitative $T_{\rm C}$ for FM ordering is generally determined from data sets as in Fig.~\ref{M-T_all_together} from the temperature of the minimum of the derivative $d\chi(T)/dT$~\cite{Cao2010, Liu2017}.  This derivative for each~$x$ is plotted in Figs.~\ref{dChi_dT_all_together}(a) and~\ref{dChi_dT_all_together}(b) for $H\parallel ab$ and $H\parallel c$, respectively.  For $x = 0.11$, a clear minimum is observed at $T_{\rm C} \approx 13$~K\@. The $T_{\rm C}$ slightly increases by $\approx 1$~K with increasing~$x$, as seen in Figs.~\ref{dChi_dT_all_together}(a) and ~\ref{dChi_dT_all_together}(b), which is only a small fraction of $T_{\rm C}(x = 0.11)$.  This very small variation of $T_{\rm C}$ with~$x$ suggests that the ferromagnetic ordering is local, i.e., arises from FM clusters of spins.

\subsection{Magnetization versus applied-magnetic-field isotherms}

\begin{figure}[h!]
\includegraphics[width = 3.4in]{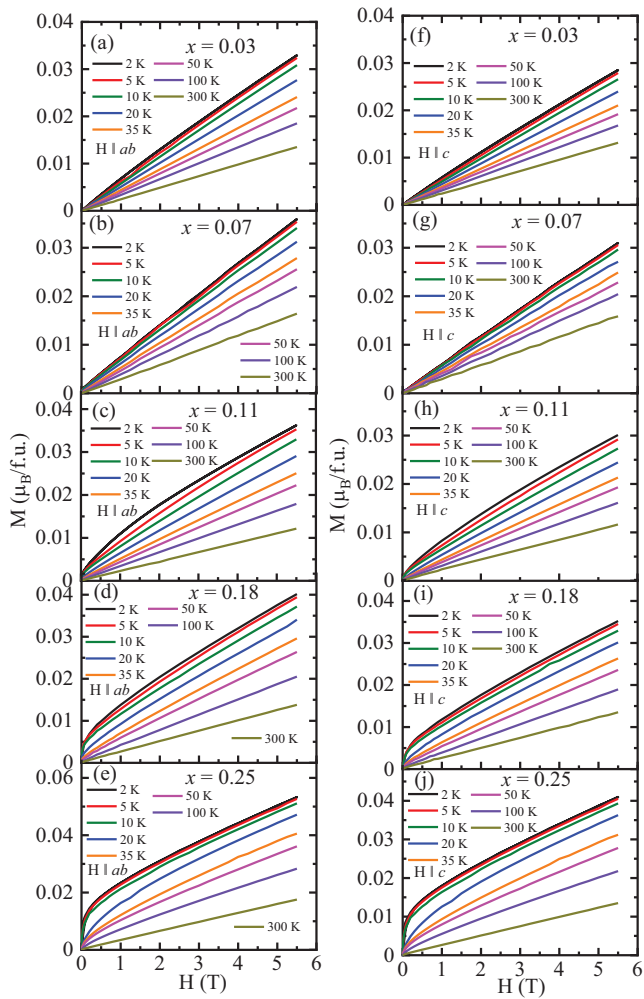}
\caption{Magnetic field $H$ dependence of the  isothermal magnetization $M$ in units of Bohr magnetons ($\mu_{\rm B}$) per formula unit (f.u.) measured at the listed temperatures for the \bcia\ crystals with (a)--(e) $H \parallel ab$ and \mbox{(f)--(j)~$H \parallel c$}.  In each panel, the temperatures of the isotherms are 2, 5, 10, 20, 35, 50, 100, and 300~K from top to bottom. }
\label{M-H_all_diff_temp}
\end{figure}

\begin{figure}
\includegraphics[width = 3.4in]{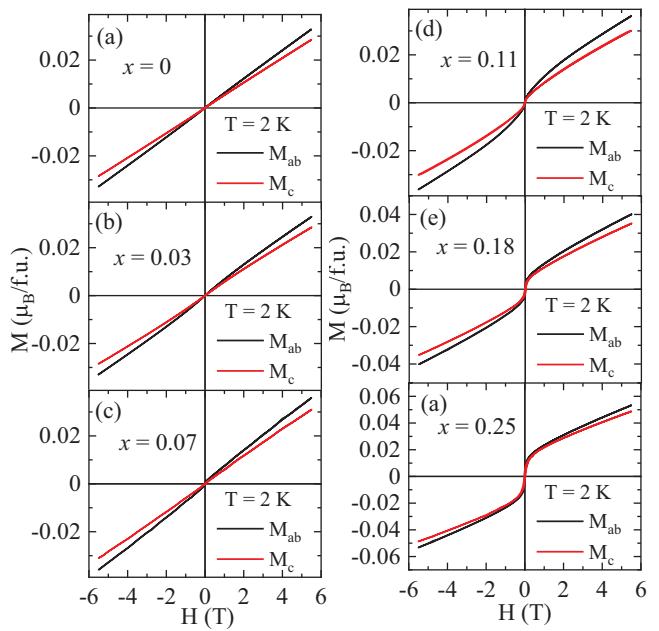}
\caption{Isothermal magnetization $M$ per formula unit vs field~$H$ measured at temperature $T = 2$~K for the \bcia\ crystals in the $ab$ plane ($M_{ab}$, black lines) and along $c$ axis ($M_{c}$, red lines).}
\label{M-H_all_separate}
\end{figure}

\begin{figure}[t]
\includegraphics[width = 3.3in]{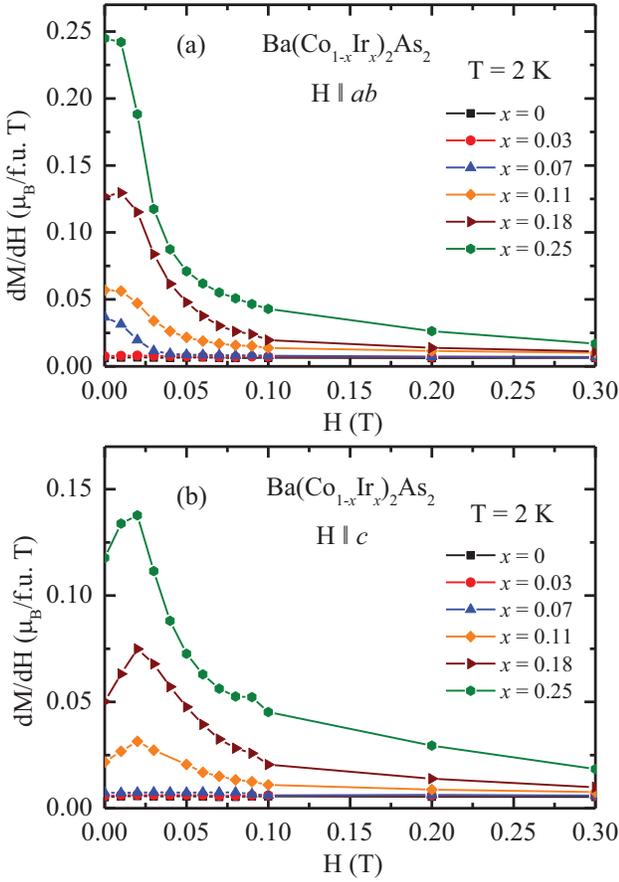}
\caption{The derivative $dM/dH$ versus $H$ at $T = 2$~K for (a)~$H \parallel ab$ and (b)~$H \parallel c$.}
\label{dM-dH_all_together}
\end{figure}

\begin{figure}[h]
\includegraphics[width = 3.4in]{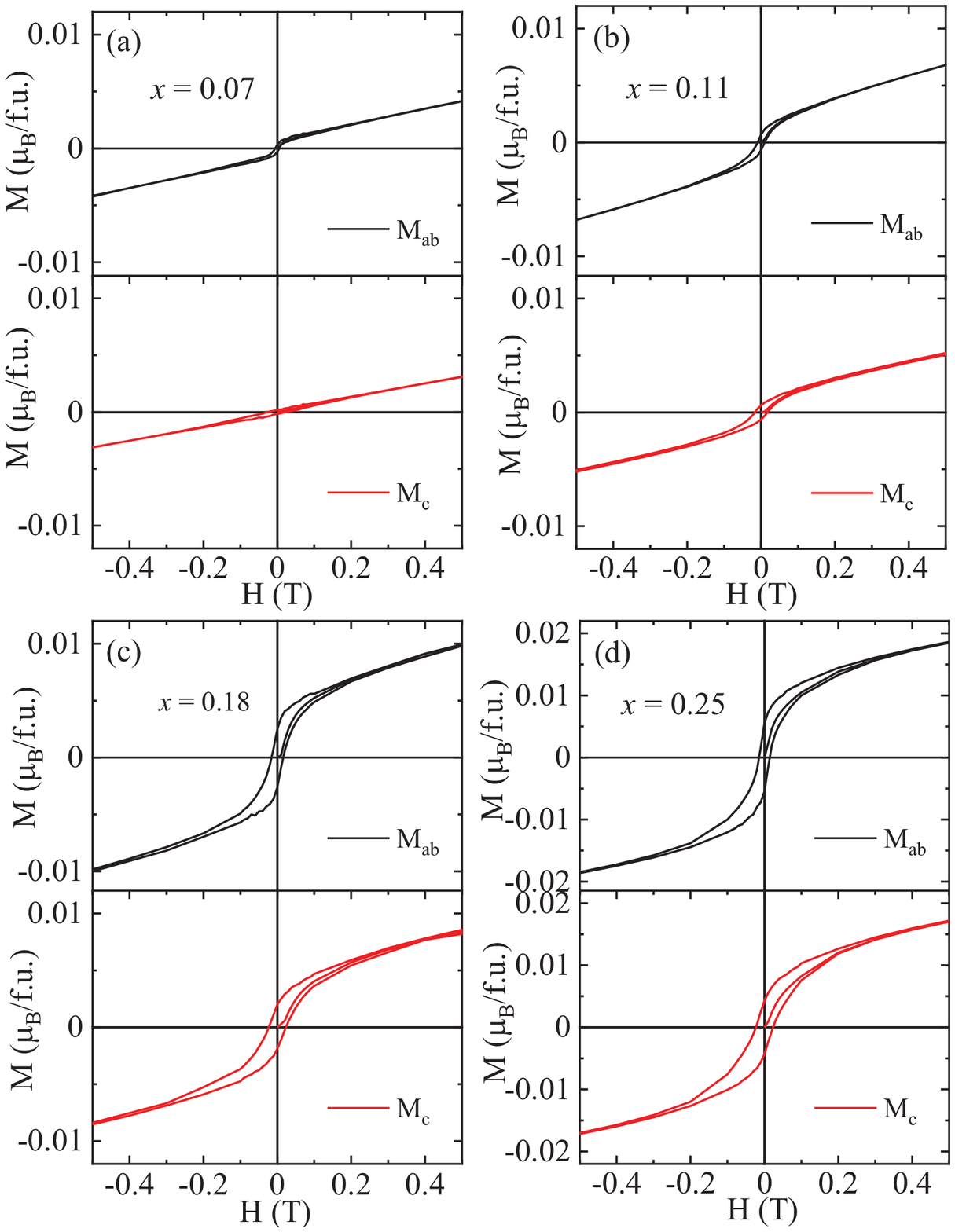}
\caption{Expanded plots of the five-quadrant magnetization~$M$ versus applied magnetic field~$H$ hysteresis isotherms at temperature \mbox{$T = 2$~K} for fields $H\parallel ab$~plane ($M_{ab}$) and $H\parallel c$~axis ($M_{c}$) for \bcia\ crystals with the Ir concentrations $x$ listed. Note the factor of two increase in the  ordinate scales in panel~(d).  }
\label{M-H_hysteresis}
\end{figure}

In order to study the field evolution of ferromagnetism in the \bcia\ crystals, $M(H)$ isotherms were measured at different temperatures in the field range \mbox{0--5.5~T} as plotted in Figs.~\ref{M-H_all_diff_temp}(a)--\ref{M-H_all_diff_temp}(e) and Figs.~\ref{M-H_all_diff_temp}(f)--\ref{M-H_all_diff_temp}(j) for $H\parallel ab$ and $H\parallel c$, respectively. For $x \geq 0.11$, the $M(H)$ isotherms are nonlinear, exhibiting negative curvature, at temperatures higher than their respective $T_{\rm C}$'s ($\approx 13$~K), suggesting the presence of dynamic FM correlations at $T>T_{\rm C}$.

Figures~\ref{M-H_all_separate}(a)--\ref{M-H_all_separate}(f) show the $M(H)$ isotherms at \mbox{$T=2$~K} for both field directions for all the studied crystals. The $M_{ab}(H)$ and $M_{c}(H)$ data for $x = 0$ and 0.03 are linear over the measured $T$ and~$H$ ranges, indicating a PM ground state. Although $M_{c}(H)$ is linear for $x = 0.07$, $M_{ab}(H)$ deviates from linearity in the low-field region with negative curvature.  This curvature is more easily visualized in the $dM_{ab}/dH$ versus~$H$ plots in Fig.~\ref{dM-dH_all_together}(a), where the corresponding data for $H\parallel c$ and $x=0.07$ in Fig.~\ref{dM-dH_all_together}(b) do not exhibit this effect. The former observation is consistent with the development of weak FM correlations in the $ab$~plane as suggested from the $\chi_{ab}(T)$ behavior in Fig.~\ref{M-T_all_together}(a) above.

Figure~\ref{M-H_hysteresis} shows expanded plots at low fields \mbox{$-0.5~{\rm T} \leq H\leq 0.5$~T} for $x=0.07$, 0.11, 0.18, and 0.25 and both field directions for each composition.   Nonlinear behavior with hysteresis is found for each composition and each field direction, with the exception of the data for $x=0.07$ which only show nonlinearity and weak hysteresis for $H\parallel ab$.  These data demonstrate behaviors typical of ferromagnets.  The remanent magnetization $M_{\rm rem}$ and coercive field $H_{\rm cf}$ of the \bcia\ crystals with $x =$ 0.11, 0.18, and 0.25 obtained from Fig.~\ref{M-H_hysteresis} are listed in Table~\ref{Tab.remcoercive}. $M_{\rm rem}$ is the same in the $ab$~plane and along the $c$~axis for $x=0.11$, but is larger in the $ab$~plane than along the $c$~axis for $x=0.18$ and 0.25.  On the other hand, $H_{\rm cf}$ is significantly larger along the $c$~axis than in the $ab$~plane for each of the crystals.  $M_{\rm rem}$ for all compositions is very small even compared to the saturation moment of $1~\mu_{\rm B}$ for spin $S=1/2$ with spectroscopic splitting factor $g=2$.

\begin{table}
\caption{\label{Tab.remcoercive} Remanent magnetization $M_{\rm rem}$ and coercive field $H_{\rm cf}$ of the \bcia\ crystals with $x = 0.11$, 0.18, and~0.25 obtained from the $M(H)$ data in Fig.~\ref{M-H_all_separate}.}
\begin{ruledtabular}
\begin{tabular}{cccc}	
Crystal								& Field direction		& $M_{\rm rem}$	 	&  $H_{\rm cf}$  \\
Composition							&				&  ($\mu_{\rm B}$/f.u.)	& (Oe) \\
\hline
Ba(Co$_{0.89}$Ir$_{0.11}$)$_2$As$_2$	& $H \parallel ab$	& 0.0007(1)	 &  93(1) \\
                                           				& $H \parallel c$	& 0.0006(1) &  168(4) \\
\hline
Ba(Co$_{0.82}$Ir$_{0.18}$)$_2$As$_2$	& $H \parallel ab$	& 0.0025(1)	 &  159(1) \\
                                           				& $H \parallel c$	& 0.0020(2)	 &  235(3) \\
 \hline
Ba(Co$_{0.75}$Ir$_{0.25}$)$_2$As$_2$	& $H \parallel ab$	& 0.0055(1)	 &  150(2) \\
                                           				& $H \parallel c$	& 0.0043(2)	 &  234(3) \\
\end{tabular}
\end{ruledtabular}
\end{table}


\section{\label{Sec:Cp} Heat-capacity measurements }

\begin{figure*}[ht!]
\includegraphics[width = 5.25in]{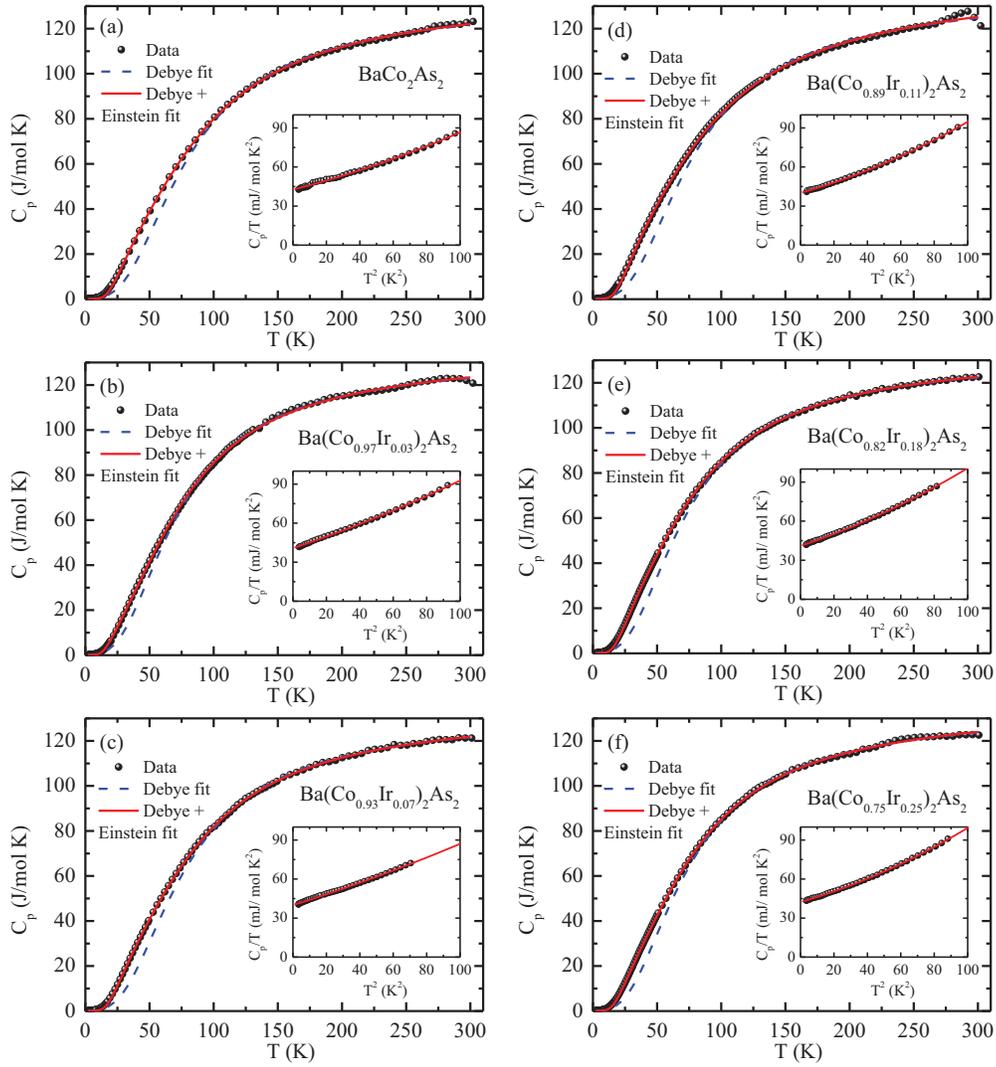}
\caption{Temperature $T$ dependences of the zero-field heat capacities $C_{\rm p}$ for the \bcia\ crystals along with the fits by the Sommerfeld and Debye models~[Eqs~(\ref{Eqs:SomDeb}), 100--300~K fits, variable $\gamma_{\rm D}$, blue dashed curves with extrapolations to $T=0$] with fit parameters given in Table~\ref{Tab.heatcapacityDebyefit}.  Also shown are fits by the sum of the Sommerfeld, Debye and Einstein terms [Eqs.~(\ref{Eqs:SomDebEin}), 25--300~K fits, variable $\gamma_{\rm D,E}$, solid red curves with extrapolations to $T=0$], with the fit parameters given in Table~\ref{Tab.heatcapacityDebyeEinsteinfit}. Insets: C$_{\rm p}$/$T$ versus $T^2$ for 1.8~K $\leq T \leq$ 10~K along with the respective fits by Eq.~(\ref{Eq.CpFit1}) with fit parameters given in Table~\ref{Tab.heatcapacitylowTfit}.}
\label{Heat_capacity_all}
\end{figure*}

\begin{table*}[ht!]
\caption{\label{Tab.heatcapacitylowTfit} Parameters obtained from the fits of heat capacity data in the temperature region 1.8~K $\leq T \leq$ 10~K\@. The listed parameters are the Sommerfeld coefficient $\gamma$, the lattice heat capacity coefficients $\beta$ and $\delta$ obtained from low-$T$ fits of C$_{\rm p}$/$T$ versus $T^2$ using Eq.~(\ref{Eq.CpFit1}), and the density of states at the Fermi energy ${\cal D}(E_{\rm F})$ derived from $\gamma$ using Eq.~(\ref{Eq:DfromGamma}) along with Debye temperature $\Theta_{\rm D}$ obtained from the value of $\beta$ using Eq.~(\ref{Eq:ThetaDfromBeta}).}
\begin{ruledtabular}
\begin{tabular}{cccccc}	
  											& $\gamma$	& $\beta$ & $\delta$  &  ${\cal D}(E_{\rm F})$ 	& $\Theta_{\rm D}$ \\
Compound 									& (mJ mol$^{-1}$ K$^{-2}$)& (mJ mol$^{-1}$ K$^{-4}$) & ($\mu$J mol$^{-1}$ K$^{-6}$)  & (states/eV\,f.u.) & (K)  \\
\hline
BaCo$_2$As$_2$                         & 42.6(2) 		& 	0.32(1)	& 	1.1(1)	&	18.07(9)		& 	312(3)	\\
Ba(Co$_{0.97}$Ir$_{0.03}$)$_2$As$_2$  & 40.7(2) 	& 	0.43(1)		& 	0.9(1)		& 	17.27(8)	&	283(2)	\\
Ba(Co$_{0.93}$Ir$_{0.07}$)$_2$As$_2$    &  	40.0(1)	&  0.40(1) 	& 	0.7(2)		&  		16.97(4)	&	290(3)	\\
Ba(Co$_{0.89}$Ir$_{0.11}$)$_2$As$_2$       	 &  40.1(2)		 & 	0.36(1)	&  	1.9(1)		& 	17.01(9)		&	300(3)	\\
Ba(Co$_{0.82}$Ir$_{0.18}$)$_2$As$_2$ 		& 	41.0(2)	&  		0.43(1)	&  1.7(2)	&  		17.39(9)		&	283(2)	\\		
Ba(Co$_{0.75}$Ir$_{0.25}$)$_2$As$_2$ 		&   42.3(2)    &  	0.37(1)	& 	2.0(1)		& 	17.95(8)		&	297(2)		\\
\end{tabular}
\end{ruledtabular}
\end{table*}

\begin{table}
\caption{\label{Tab.heatcapacityDebyefit} Sommerfeld coefficient $\gamma_{\rm D}$ and Debye temperature $\Theta_{\rm D}$ determined by fitting the $C_{\rm p}(T)$ data for \bcia\ crystals in the temperature range \mbox{100~K $\leq T \leq$ 300~K} by Eqs.~(\ref{Eqs:SomDeb}). The fixed $\gamma_{\rm D}$ for BaCo$_2$As$_2$ is the low-$T$ $\gamma$ value for this compound  from Table~\ref{Tab.heatcapacitylowTfit}.}
\begin{ruledtabular}
\begin{tabular}{ccc}	
  											&  $\gamma_{\rm D}$ 	& $\Theta_{\rm D}$ \\
Compound 									&$\left(\rm{\frac{mJ}{mol\, K^2}}\right)$ & (K)  \\
\hline
BaCo$_2$As$_2$                         & 	42.6 (fixed)		& 	368(6)	\\
                                   & 	13(1)		& 	325(2)	\\
Ba(Co$_{0.97}$Ir$_{0.03}$)$_2$As$_2$  &  	13.5(7)	&	293(2)	\\
Ba(Co$_{0.93}$Ir$_{0.07}$)$_2$As$_2$    &  		11.3(4)	&	315(1)	\\
Ba(Co$_{0.89}$Ir$_{0.11}$)$_2$As$_2$       	 &  	22(1)		&	317(2)	\\
Ba(Co$_{0.82}$Ir$_{0.18}$)$_2$As$_2$ 		& 		11.9(4)		&	299(1)	\\		
Ba(Co$_{0.75}$Ir$_{0.25}$)$_2$As$_2$ 		&   	16.7(1)		&	299(1)		\\
\end{tabular}
\end{ruledtabular}
\end{table}

\begin{table}
\caption{\label{Tab.heatcapacityDebyeEinsteinfit} Sommerfeld coefficient $\gamma_{\rm D,E}$,  Einstein contribution~$\alpha$ to the heat capacity, Einstein temperature $\Theta_{\rm E}$, and the Debye temperature $\Theta_{\rm D}$ obtained by fitting the $C_{\rm p}(T)$ data for \bcia\ crystals in the temperature range \mbox{25--300~K} by Eqs.~(\ref{Eq:DebyeTerm}) and~(\ref{Eqs:SomDebEin}).}
\begin{ruledtabular}
\begin{tabular}{ccccc}	
  											& $\gamma_{\rm D,E}$	& $\alpha$ & $\Theta_{\rm E}$ 	& $\Theta_{\rm D}$ \\
Compound 									& $\left(\rm{\frac{mJ}{mol\, K^2}}\right)$&  & (K) & (K)  \\
\hline
BaCo$_2$As$_2$                         & 42.6 (fixed) 		& 	0.67(2)	& 		161(4)		& 	704(35)	\\
                                       & 15.9(4) 		& 	0.30(1)	& 		111(2)		& 	396(3)	\\
Ba(Co$_{0.97}$Ir$_{0.03}$)$_2$As$_2$  & 16.1(7) 	& 	0.27(2)		& 		98(4)	&	340(4)	\\
Ba(Co$_{0.93}$Ir$_{0.07}$)$_2$As$_2$    &  	14.3(3)	&  0.35(1) 	& 	  116(1)	&	395(3)	\\
Ba(Co$_{0.89}$Ir$_{0.11}$)$_2$As$_2$       	 &  23.7(7)		 & 	0.39(1)	&  	 	119(2)		&	416(6)	\\
Ba(Co$_{0.82}$Ir$_{0.18}$)$_2$As$_2$ 		& 	14.5(2)	&  		0.35(1)	&  	110(1)		&	373(2)	\\		
Ba(Co$_{0.75}$Ir$_{0.25}$)$_2$As$_2$ 		&   19.1(4)    &  	0.30(1)	& 	107(2)		&	361(3)		\\
\end{tabular}
\end{ruledtabular}
\end{table}

\begin{figure}
\includegraphics[width = 3in]{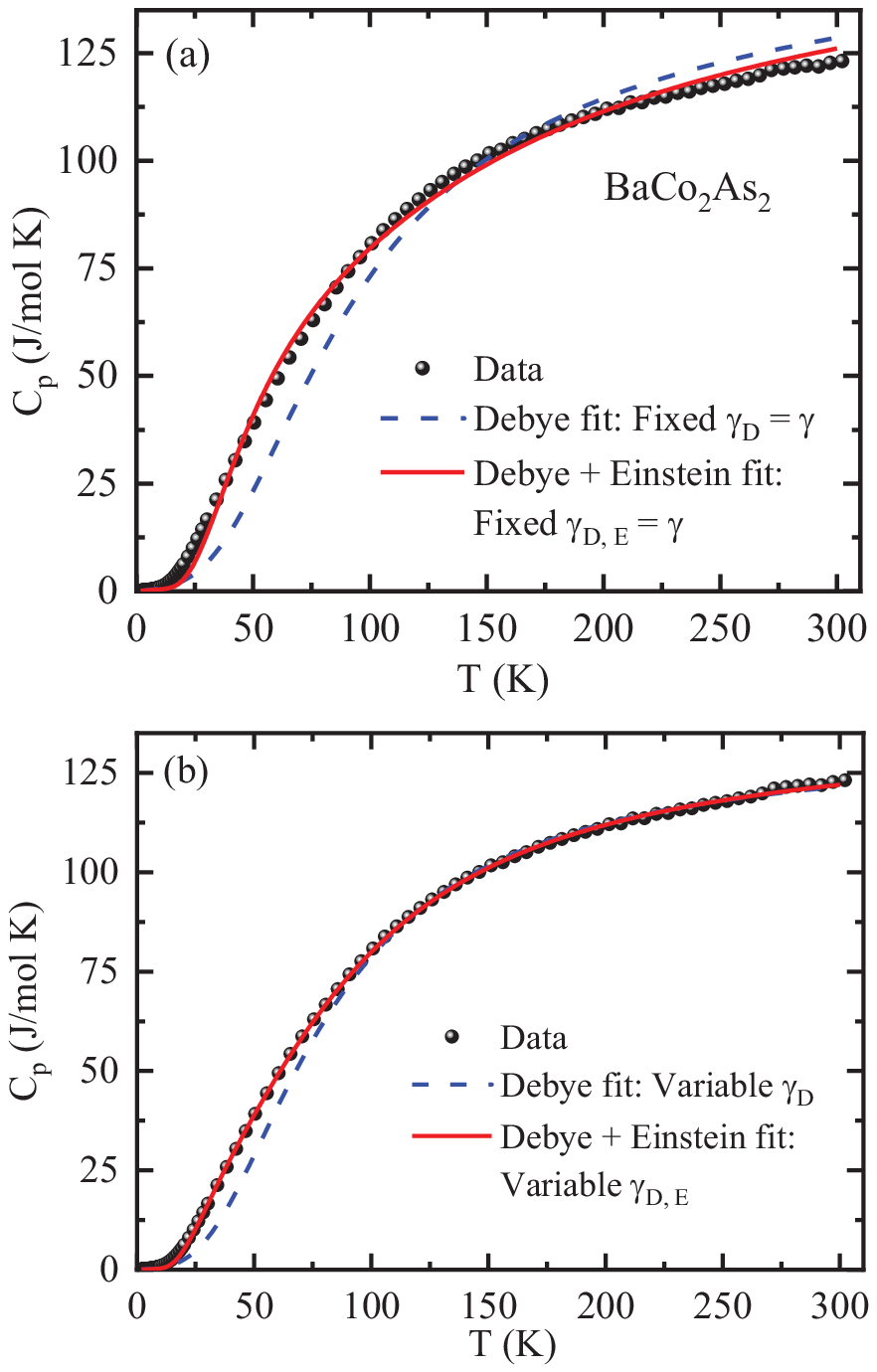}
\caption{(a) and (b)~Temperature~$T$ dependence of the zero-field heat capacity $C_{\rm p}$ for a \bca\ crystal from 2 to 300~K\@. Panel~(a) shows a fit from 100 to 300~K by the Sommerfeld plus Debye models in Eqs.~(\ref{Eqs:SomDeb}) (blue dashed line, with an extrapolation to $T=0$), where $\gamma_{\rm D}$ was taken to be the fixed low-$T$ $\gamma$ value in Table~\ref{Tab.heatcapacitylowTfit}.  The solid red curve is a fit by the sum of the Summerfeld, Debye and Einstein terms over the temperature range 25--300~K with an extrapolation to $T=0$ where $\gamma_{\rm D}$ was again fixed to the low-$T$ $\gamma$~value. Panel~(b) shows a fit from 100 to 300~K by the Sommerfeld plus Debye models in Eqs.~(\ref{Eqs:SomDeb}) (blue dashed line, with an extrapolation to $T=0$), where $\gamma_{\rm D}$ was a variable in the fit.  The solid red curve is a fit by the sum of the Summerfeld, Debye and Einstein terms over the temperature range 25--300~K with an extrapolation to $T=0$ where $\gamma_{\rm D}$ was again a variable in the fit. }
\label{Heat_capacity_BaCo2As2}
\end{figure}

The zero-field heat capacities versus temperature $C_{\rm p}(T)$ of the \bcia\ crystals were measured in the temperature range 1.8--300~K and are plotted in Figs.~\ref{Heat_capacity_all}(a)--\ref{Heat_capacity_all}(f).  No signature of any transition is observed for any of the crystals in the measured temperature range.  This null result may be associated with the low ordered moments, the suggested short-range nature of the FM order, and/or with an itinerant nature of the ferromagnetism for which calorimetric features at $T_{\rm C}$ are difficult to detect. For each crystal, $C_{\rm p}(T)$ approaches the classical Dulong-Petit limit $C_{\rm p} = 3nR = 124.7$~J/mol\,K at 300~K where $R$ is the molar gas constant and $n = 5$ is the number of atoms per formula unit.

The low-temperature $C_{\rm p}(T)$ data are described well by
\bea
C_{\rm p}(T) = \gamma T+ \beta T^3 +\delta T^5,
\label{Eq.CpFit1}
\eea
where $\gamma$ is the Sommerfeld coefficient associated with degenerate itinerant current carriers, and $\beta$ and $\delta$ are coefficients associated with the low-$T$ lattice contribution to the  heat capacity.  The insets of \mbox{Figs.~\ref{Heat_capacity_all}(a)--\ref{Heat_capacity_all}(f)} show $C_{\rm p}(T)/T$ versus $T^2$ in the temperature region 1.8--10~K, along with fits by Eq.~(\ref{Eq.CpFit1}). The fitted parameters are listed in Table~\ref{Tab.heatcapacitylowTfit}. The value \mbox{$\gamma = 42.6(2)$~mJ/mol\,K$^{2}$} obtained for \bca\ is similar to the values  reported earlier~\cite{Sefat2009,Anand2014}. No significant change in the $\gamma$ values is observed in the Ir-substituted compounds. This result contrasts with the calorimetric properties of \ccia\ crystals where $\gamma$ significantly increases with increasing Ir concentration~\cite{Pakhira2020ccia}.

The electronic densities of states at the Fermi energy ${\cal D}(E_{\rm F})$ for the \bcia\ crystals were obtained from the fitted $\gamma$ values using the relation
\bea
{\cal D}_\gamma(E_{\rm F})\rm  {\left[  \frac{states}{eV\,f.u.} \right] = \frac{1}{2.357}\ \gamma \left[\frac{mJ}{mol\,K^2}\right]   },
\label{Eq:DfromGamma}
\eea
where this expression includes the factor of two Zeeman degeneracy of the conduction carriers. The derived values of ${\cal D}_\gamma(E_{\rm F})$ for the crystals studied here are listed in Table~\ref{Tab.heatcapacitylowTfit}.  The Debye temperature $\Theta_{\rm D}$ is obtained from $\beta$ using the relation
\bea
\Theta_{\rm D} = \left(\frac{12\pi^4nR}{5\beta}\right)^{1/3}.
\label{Eq:ThetaDfromBeta}
\eea
The $\Theta_{\rm D}$ values obtained from the respective $\beta$ values in Table~\ref{Tab.heatcapacitylowTfit} for our \bcia\ crystals are also listed in Table~\ref{Tab.heatcapacitylowTfit}.

The $C_{\rm p}(T)$ data in Fig.~\ref{Heat_capacity_all} were initially fitted over larger temperature ranges by the Sommerfeld electronic heat capacity plus the Debye model for the lattice heat capacity using the relations
\bse
\label{Eqs:SomDeb}
\bea
C_{\rm p}(T) &=& \gamma_{\rm D} T+ nC_{\rm V\,Debye}(T),\label{Eq:Debye_Fit} \\*
\hspace{-1in} C_{\rm V\,Debye}(T) &=& 9R \left(\frac{T}{\Theta_{\rm D}}\right)^3\int_{0}^{\Theta_{\rm D}/T}\frac{x^4e^x}{(e^x-1)^2} dx,\label{Eq:DebyeTerm}
\eea
\ese
where $\gamma_{\rm D}$ is the Sommerfeld coefficient and $R$ is the molar-gas constant. We used the Pad\'e approximant for the Debye function given in Ref.~\cite{Goetsch_2012} for calculating the lattice heat capacity. The $C_{\rm p}(T)$ data in the fitted temperature range 100~K~$\leq T\leq$~300~K are not described well by Eq.~(\ref{Eq:Debye_Fit}) using the fixed $\gamma$ values from Table~\ref{Tab.heatcapacitylowTfit} for $\gamma_{\rm D}$ as illustrated by the dashed blue curve for \bca\ in Fig.~\ref{Heat_capacity_BaCo2As2}(a) with an extrapolation to $T=0$.  Similarly-bad fits were obtained for the Ir-substituted  crystals using the respective fixed $\gamma$ values (not shown).

Therefore we took $\gamma_{\rm D}$ to be a variable fitting parameter in Eq.~(\ref{Eq:Debye_Fit}).  Now the $C_{\rm p}(T)$ data for \bca\ are fitted better for $T \gtrsim 100$~K, as depicted by the blue dashed curve in Fig.~\ref{Heat_capacity_BaCo2As2}(b) with an extrapolation to $T=0$.  The corresponding blue-dashed fitted curves are shown in Figs.~\ref{Heat_capacity_all}(a)--\ref{Heat_capacity_all}(f) for all the \bcia\ crystals. The fitted $\gamma_{\rm D}$ values in Table~\ref{Tab.heatcapacityDebyefit} differ significantly from the $\gamma$ values in Table~\ref{Tab.heatcapacitylowTfit} obtained from the low-$T$ $C_{\rm p}(T)$ fits. However, the obtained $\Theta_{\rm D}$ values in Table~\ref{Tab.heatcapacityDebyefit} are similar to those in Table~\ref{Tab.heatcapacitylowTfit} obtained from the $\beta$ values in the low-$T$ fits.

In Figs.~\ref{Heat_capacity_all} and~\ref{Heat_capacity_BaCo2As2}(b), the 100--300~K fits by Eqs.~(\ref{Eqs:SomDeb}) with a fitted $\gamma_{\rm D}$ are quite good, but the extrapolations to lower temperatures are seen to be poor agreement with the corresponding data for all compositions.  Therefore in order to fit the heat capacity over a larger temperature range, we added an Einstein lattice heat capacity contribution to Eqs.~(\ref{Eqs:SomDeb}), yielding
\bse
\label{Eqs:SomDebEin}
\bea
\label{Eq:Debye and Einstein}
C_{\rm p}(T) = \gamma_{\rm {D,E}} T + (1 - \alpha)C_{\rm V\,Debye} + \alpha C_{\rm V\,Einstein},
\eea
where
\bea
\label{Eq:Einstein}
C_{\rm V\,Einstein}(T) = 3R \left(\frac{\Theta_{\rm E}}{T}\right)^2\frac{e^{\Theta_{\rm E}/T}}{(e^{\Theta_{\rm E}/T} - 1)^2},
\eea
\ese
$\gamma_{\rm {D,E}}$ is the Sommerfeld coefficient associated with this fit and is a fitting parameter, $\alpha$ is the fraction of the Einstein contribution to the total lattice heat capacity, and $\Theta_{\rm E}$ is the Einstein temperature.  The red solid curves in Figs.~\ref{Heat_capacity_all}(a)--\ref{Heat_capacity_all}(f) and \ref{Heat_capacity_BaCo2As2}(b) are the fits of the $C_{\rm p}(T)$ data from 25 to 300~K by Eqs.~(\ref{Eqs:SomDebEin}) and the fitted parameters are listed in Table~\ref{Tab.heatcapacityDebyeEinsteinfit}.  The fits are now seen to be very good from 25 to 300~K\@.  This analysis suggests that there exist low-frequency optic modes associated with the heavy Ba atoms in the \bcia\ crystals.


\section{\label{Sec:Summary} Concluding Remarks}

Single-crystalline samples of the Ir-substituted cobalt arsenides \bcia\ ($0\leq x\leq 0.25$) were grown from self-flux and characterized by means of single-crystal x-ray diffraction.  Anisotropic magnetic susceptibility~$\chi$, field-dependent of isothermal magnetization, and heat capacity $C_{\rm p}$ measurements were carried out versus temperature~$T$\@. The room-temperature single-crystal structural analyses reveal that all the studied compositions form in an uncollapsed body-centered-tetragonal ${\rm ThCr_2Si_2}$ structure with space group $I4/mmm$.  We could not obtain homogeneous crystals for $x > 0.25$ with our self-flux crystal-growth technique. 

Although no magnetic ordering was observed for undoped \bca\ as previously reported, the $\chi$ at low~$T$ increases significantly with increasing Ir substitution suggesting the presence of ferromagnetic (FM) interactions.  A clear signature of FM ordering was observed in the magnetic susceptibility $\chi(T)$ data with a nearly composition-independent Curie temperature $T_{\rm C} \approx 13$~K for $x = 0.11$--0.25. From the low-$T$ bifurcation between the ZFC and FC susceptibilities, invariance of $T_{\rm C}$ with $x$ for $0.11\leq x\leq
0.25$, and the very small spontaneous and remanent magnetizations, the FM state below 13~K is likely associated with short-range FM ordering of spin clusters.

The $\chi_{ab}$ and $\chi_c$ are generally anisotropic for our crystals in the measured $T$ range with $\chi_{ab} > \chi_c$. The zero-field $C_{\rm p}(T)$ data for the \bcia\ crystals do not show any feature associated with the magnetism in the temperature range 1.8--300~K, consistent with the inferred short-range FM ordering of spin clusters with very small spontaneous moments.  The Sommerfeld coefficient $\gamma$ obtained from low-$T$ heat-capacity measurements is found to be similar for all the studied compositions. The $C_{\rm p}(T)$ data over a larger temperature range in the paramagnetic region could be described by including both Debye and Einstein lattice heat capacity contributions.

Although our study does not directly confirm that \bca\ is near a FM quantum-critical point~\cite{Sefat2009},  we do  find that Ir substitutions for Co in \bca\ result in short-range FM ordering.

\acknowledgments

This research was supported by the U.S. Department of Energy, Office of Basic Energy Sciences, Division of Materials Sciences and Engineering.  Ames Laboratory is operated for the U.S. Department of Energy by Iowa State University under Contract No.~DE-AC02-07CH11358.


\end{document}